\documentclass[sigconf,screen=true]{acmart}

\usepackage[ruled,vlined,lined,commentsnumbered]{algorithm2e}
\usepackage{amsmath,amsfonts}
\usepackage{array}
\usepackage{booktabs}
\usepackage[skip=1pt,labelfont=bf]{caption}
 \usepackage{calligra}
\usepackage{color, colortbl}
\usepackage{courier}
\usepackage{csvsimple}
\usepackage{enumitem}
\usepackage{fancybox}
\usepackage{fontenc}
\usepackage{graphicx}
\usepackage{listings}
\usepackage{longtable}
\usepackage{lscape}
\usepackage{makecell}
\usepackage{marvosym}
\usepackage{moreverb}
\usepackage{multicol}
\usepackage{multirow}
\usepackage{pifont}
\usepackage{rotating}
\usepackage{setspace}
\usepackage{caption}
\usepackage{subfigure}
\usepackage[most]{tcolorbox}
\usepackage{threeparttable}
\usepackage{tikz}
\usepackage[normalem]{ulem}
\usepackage{url}
\usepackage{soul}
\usepackage{wasysym}
\usepackage{svg}
\usepackage{textcomp}
\usepackage{xcolor}
\usepackage{wrapfig}
\usepackage{pdflscape}
\usepackage{hyperref}

\usepackage{amsmath}
\usepackage{array}
\usepackage{balance}
\usepackage{microtype}
\usepackage{microtype}

\newcolumntype{L}[1]{>{\raggedright\let\newline\\\arraybackslash\hspace{0pt}}m{#1}}
\newcolumntype{C}[1]{>{\centering\let\newline\\\arraybackslash\hspace{0pt}}m{#1}}
\newcolumntype{R}[1]{>{\raggedleft\let\newline\\\arraybackslash\hspace{0pt}}m{#1}}

\newboolean{showcomments}
\setboolean{showcomments}{true}
\ifthenelse{\boolean{showcomments}}
 { \newcommand{\mynote}[2]{
      \fbox{\bfseries\sffamily\scriptsize#1}
        {\small$\blacktriangleright$\textsf{\emph{#2}}$\blacktriangleleft$}}}
        { \newcommand{\mynote}[2]{}}

\newcommand{\xy}[1]{\textcolor{black}{#1}}

\newcommand{\toolname}{ThinkRepair\xspace}

\definecolor{darkgreen}{rgb}{114,169,119}

\newcommand{\intuition}[1]{
\begin{tcolorbox}[colback=white,boxrule=1pt,top=0pt,bottom=0pt,left=1pt,right=2pt,top=2pt,bottom=2pt]
\em #1
\end{tcolorbox}
}

\AtBeginDocument{%
  \providecommand\BibTeX{{%
    \normalfont B\kern-0.5em{\scshape i\kern-0.25em b}\kern-0.8em\TeX}}}

\setcopyright{acmlicensed}
\acmDOI{10.1145/3650212.3680359}
\acmYear{2024}
\copyrightyear{2024}
\acmISBN{979-8-4007-0612-7/24/09}
\acmConference[ISSTA '24]{Proceedings of the 33rd ACM SIGSOFT International Symposium on Software Testing and Analysis}{September 16--20, 2024}{Vienna, Austria}
\acmBooktitle{Proceedings of the 33rd ACM SIGSOFT International Symposium on Software Testing and Analysis (ISSTA '24), September 16--20, 2024, Vienna, Austria}
\acmSubmissionID{issta24main-p840-p}
\received{2024-04-12}
\received[accepted]{2024-07-03}

\begin{document}

\title{\toolname: Self-Directed Automated Program Repair}







\author{Xin Yin}
\orcid{0009-0005-3396-0571}
\affiliation{%
  \institution{The State Key Laboratory of Blockchain and Data Security, Zhejiang University}
  \city{Hangzhou}
  \country{China}
}
\email{xyin@zju.edu.cn}

\author{Chao Ni}
\orcid{0000-0002-2906-0598}
\authornote{Chao Ni is the corresponding author.\\
He is also with Hangzhou High-Tech Zone (Binjiang) Institute of Blockchain and Data Security.}
\affiliation{%
  \institution{The State Key Laboratory of Blockchain and Data Security, Zhejiang University}
  \city{Hangzhou}
  \country{China}
}
\email{chaoni@zju.edu.cn}

\author{Shaohua Wang}
\orcid{0000-0001-5777-7759}
\affiliation{%
  \institution{Central University of Finance and Economics}
  \city{Beijing}
  \country{China}
}
\email{davidshwang@ieee.org}

\author{Zhenhao Li}
\orcid{0000-0002-4909-1535}
\affiliation{%
  \institution{Concordia University}
  \city{Toronto}
  \country{Canada}
}
\email{lzh9410@gmail.com}

\author{Limin Zeng}
\orcid{0000-0003-1009-7603}
\affiliation{%
  \institution{Zhejiang University}
  \city{Hangzhou}
  \country{China}
}
\email{limin.zeng@zju.edu.cn}

\author{Xiaohu Yang}
\orcid{0000-0003-4111-4189}
\affiliation{%
  \institution{Zhejiang University}
  \city{Hangzhou}
  \country{China}
}
\email{yangxh@zju.edu.cn}

\begin{abstract}
Though many approaches have been proposed for Automated Program Repair (APR) and indeed achieved remarkable performance, they still have limitations in fixing bugs that require analyzing and reasoning about the logic of the buggy program.
Recently, large language models (LLMs) instructed by prompt engineering have attracted much attention for their powerful ability to address many kinds of tasks including bug-fixing.
However, the quality of the prompt will highly affect the ability of LLMs and manually constructing high-quality prompts is a costly endeavor.

To address this limitation, we propose a self-directed LLM-based automated program repair, \toolname, with two main phases: collection phase and fixing phase.
The former phase automatically collects various chains of thoughts that constitute pre-fixed knowledge by instructing LLMs with the Chain-of-Thought (CoT) prompt.
The latter phase targets fixing a bug by first selecting examples for few-shot learning and second automatically interacting with LLMs, optionally appending with feedback of testing information.

Evaluations on two widely studied datasets (Defects4J and Quix-Bugs) by comparing \toolname with 12 SOTA APRs indicate the priority of \toolname in fixing bugs.
Notably, \toolname fixes 98 bugs and improves baselines by 27\%$\sim$344.4\% on Defects4J V1.2. 
On Defects4J V2.0, \toolname fixes 12$\sim$65 more bugs than the SOTA APRs. 
Additionally, \toolname also makes a considerable improvement on QuixBugs (31 for Java and 21 for Python at most).
\end{abstract}



\maketitle

\vspace{-0.3cm}
\section{Introduction}

Automated Program Repair (APR) is a promising approach to automatically fix bugs in computer programs, which can significantly reduce debugging time and enhance software reliability. 
Traditional APR techniques can be classified into heuristic-based~\cite{le2016history, le2011genprog, wen2018context}, constraint-based~\cite{demarco2014automatic, le2017s3, long2015staged, mechtaev2016angelix}, and template-based~\cite{liu2019tbar, ghanbari2019practical, hua2018sketchfix, liu2019avatar, martinez2016astor} approaches.
Template-based APRs can fix a large number of bugs using predefined templates, but are limited to these patterns and lack generalizability to other types of bugs.
To address this limitation, techniques based on Neural Machine Translation (NMT) have been extensively studied in recent years~\cite{jiang2023knod, meng2023tenure, ye2022selfapr, ye2022neural, zhu2021syntax, jiang2021cure, drain2021deepdebug}.
These approaches treat fixing bugs as an NMT problem, where the goal is to translate buggy code into correct code, rely heavily on bug-fixing datasets obtained from open-source repositories.

To overcome the limitations of NMT-based APR, researchers are exploring the use of pre-trained LLMs, which generate correct code directly based on context, mitigating the need for translation from buggy code by pre-training on large amounts of open-source code snippets.
AlphaRepair~\cite{xia2022less} is the first tool for cloze-style APR and its performance indicates that LLM-based APR outperforms the widely studied NMT-based APR techniques. 
Following that, researchers~\cite{prenner2022can, kolak2022patch} adopt Codex to generate a repaired code function based on the buggy one.
Recently, Xia et al.~\cite{xia2023automated} conducted an extensive study of LLM-based APR using various LLMs~\cite{chen2021evaluating, black2022gpt, wang2021codet5, fried2022incoder} by In-Context Learning, further demonstrated the superiority of LLM-based APR.

Instead of directly generating an answer, Chain-of-Thought (CoT) prompting~\cite{wei2022chain} instructs LLMs to obtain an answer with a step-by-step process, which largely improves performance on reasoning.
Despite the success of CoT, studies~\cite{liu2021makes, lu2021fantastically} have shown that the strength of LLMs depending on the few-shot examples, and low-quality examples are unable to guide LLMs to engage more profound inferential reasoning.

In this paper, we aim to advance APR by introducing \toolname, an approach with strong analyzing and reasoning capabilities for program repair tasks. 
\xy
{
\textbf{First, we propose an LLM-based framework for APR.} 
LLMs are trained in an unsupervised fashion using up to billions of text/code tokens.
This large-scale unsupervised learning process allows LLMs to have strong reasoning thinking and can be applied for program repair without relying on historical bug fixes.
Therefore, we propose a novel LLM-based approach \toolname for APR since representative conversational LLM model provides advanced capabilities for several tasks, including natural language processing~\cite{openai2022chatgpt}, code generation~\cite{li2023starcoder}, and bug-fixing~\cite{sobania2023analysis, haque2023potential}.
\textbf{Second, we develop a self-directed framework to enhance the LLM's capabilities.}
Fixing bugs requires logical thinking and a coherent series of intermediate steps and few-shot CoT enables LLMs to improve their analytical and reasoning capabilities through a step-by-step process instead of generating fixed code directly.
However, previous studies~\cite{white2023prompt, liu2023improving, li2023mot} have shown that the quality of prompts will highly affect LLM's reasoning abilities across various tasks.
Meanwhile, manually constructing high-quality prompts is a costly endeavor.
Therefore, we propose \toolname, which contains two phases: (1) the collection phase aims to construct chains of thoughts that constitute pre-fixed knowledge pool and (2) the fixing phase aims to fix a buggy function by selecting high-quality examples from pre-fixed knowledge pool for few-shot learning and interacting with LLM with optional feedback testing information. 
}

We conduct experiments on two widely studied dataset (i.e., Defects4J~\cite{just2014defects4j} and QuixBugs~\cite{lin2017quixbugs}) by comparing \toolname with 12 state-of-the-art APR approaches.
The results indicate the priority of \toolname over baselines.
For example, on Defects4J V1.2, \toolname totally fixes 98 bugs and improves baselines by 27\%$\sim$344.4\%.
\toolname also achieve the best performance on Defects4J V2.0 and fixes 12$\sim$65 more bugs than SOTAs.
Our results also show that \toolname has a complementary results to the SOTAs and exclusively fixes 32 bugs (out of 98) that the SOTAs can not fix.
\xy
{
We also collect bugs from real-world projects to evaluate data leakage.
\toolname can fix 19 out of 44 bugs on RWB V1.0, and 10 out of 29 bugs on RWB V2.0.
}

In summary, the key contributions of this paper include:

\textbf{A. Novel Self-directed LLM-based APR:} \toolname advances LLM-based APR for program bugs.
We show that LLM-based APR can achieve comparable and complementary results as other APR directions.

\textbf{B. Automatic Reasoning for APR:}
(1) Few-shot CoT that largely enhances analyzing and reasoning capabilities to understand the semantics of the buggy function;
(2) The framework with automated chains of thoughts collection, few-shot selection and interaction feedback to promote reasoning for APR.

\textbf{C. Extensive Evaluation:} 
\xy
{
(1) We evaluate \toolname against current state-of-the-art NMT-based and LLM-based tools on the widely
studied Defects4J~\cite{just2014defects4j} and QuixBugs~\cite{lin2017quixbugs} datasets;
(2) We also conduct a further study about the data leakage in \toolname by collecting new datasets from real-world projects.
}
\label{sec:introduction}

\vspace{-0.3cm}
\section{Motivation}

\subsection{A Motivation Example}

\begin{figure}[htbp]
    \centering
    \vspace{-0.2cm}
    \includegraphics[width=.55\linewidth]{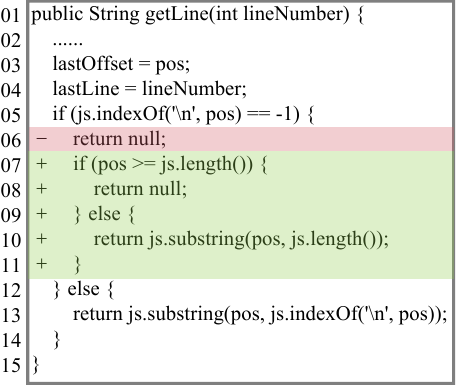}
    \caption{\href{https://storage.googleapis.com/google-code-archive/v2/code.google.com/closure-compiler/issues/issue-538.json}{Closure-56}: a code logic error in Closure project}
    \vspace{-0.3cm}
    \label{fig:motivation}
\end{figure}

Fig.~\ref{fig:motivation} shows the bug (\href{https://storage.googleapis.com/google-code-archive/v2/code.google.com/closure-compiler/issues/issue-538.json}{Closure-56}) and its fix of a Java project named Closure. 
The function's purpose is to extract a specified line of text from the given text content.
It takes a \textit{``lineNumber''} as input and returns the content of the corresponding line. 
The function first retrieves the text content and searches line by line from the beginning until it finds the specified line number or reaches the end of the file. 
If it successfully finds the specified line, it returns the content of that line. 
However, there is a logical error inside the function (i.e., Line 6). 
When it fails to retrieve the text content or the specified line number is invalid, this buggy function incorrectly returns \textit{``null''}. 
To fix this bug, a developer modified the return statement inside the \textit{``if (js.indexOf(`\textbackslash{}n', pos) == -1)''} block. 
The updated code snippet resolves the bug by checking if the variable \textit{``pos''} has exceeded the length of the \textit{``js''} string. 
If it does, it returns \textit{``null''}, indicating that the requested line number is out of bounds. 
Otherwise, it returns the \textit{``substring''} from \textit{``pos''} to the end of the \textit{``js''} string, effectively returning the content of the requested line.

To fix the logic error in Fig.~\ref{fig:motivation}, one needs to understand the semantics of the function. 
For example, the purpose of \textit{``getLine''} is to extract a specified line of text from the given text content, with two additional handling cases when it cannot find the next line break. 
Fixing this bug requires to add additional code, which is more challenging than modifying or deleting code to fix a bug, as it demands a greater understanding ability (e.g., considering new code boundaries or new code functionality).

\begin{figure*}[htbp]
    \centering
    \vspace{-0.2cm}
    \includegraphics[width=.75\linewidth]{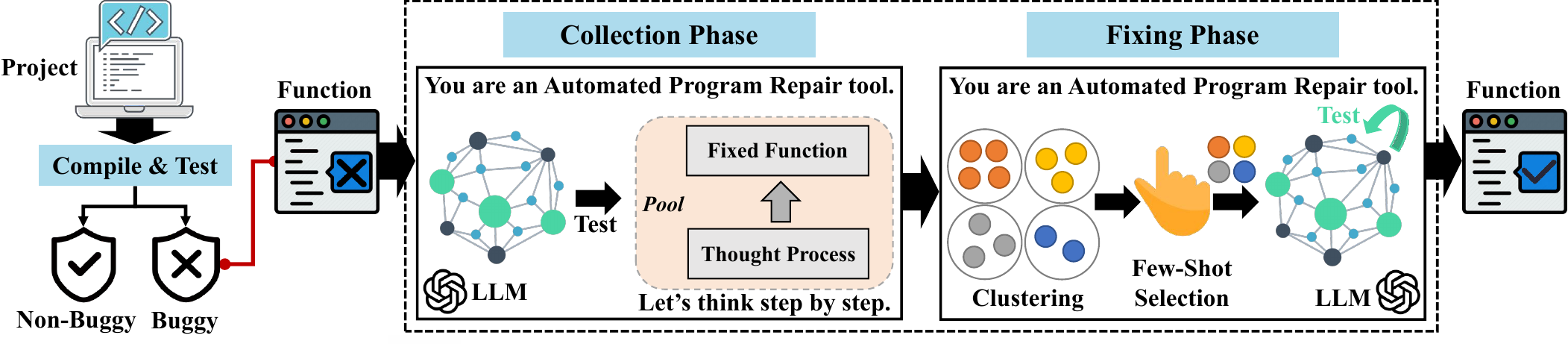}
    \caption{Overview of \toolname}
    \label{fig:overview}
    \vspace{-0.3cm}
\end{figure*}

\textbf{Observation 1.} 
\ul{Fixing the above bug requires powerful code understanding and reasoning about the code logic for the given buggy function.}
Over the years, several NMT-based APRs~\cite{jiang2023knod, meng2023tenure, ye2022selfapr, ye2022neural, jiang2021cure} have been proposed, and show strong bug fixing capabilities through training on large amounts of labeled data. However, none of them has possessed powerful analytical reasoning capabilities to auto-fix the above bug, such as KNOD~\cite{jiang2023knod} and SelfAPR~\cite{ye2022selfapr}. If there are no similar repair patterns in their training data, it becomes difficult to correctly fix the issue, as none of them can understand and reason to add new logic into the code for fixing.

\textbf{Observation 2.} 
\ul{Powerful models need to be empowered and guided with prior fix knowledge.}
Unlike current NMT-based APRs using limited number of bug fixes as training data, LLM is directly pre-trained using millions of code snippets from open-source projects, allowing it to provide a variety of edit types to fix different bugs. 
LLM has shown dominantly superior reasoning capabilities over any other existing AI models in natural language and code understanding~\cite{bang2023multitask}.
We observed that LLM can correctly understand the code in Fig.~\ref{fig:motivation}, but cannot auto-fix it due to the missing reasoning required by fixing.
The performance of LLM is influenced by prompts, and low-quality prompts are unable to guide LLM to engage in more profound inferential reasoning to fix such bugs.

\subsection{Key Ideas}

Based on the above observations, we propose an LLM-based APR framework using chain-of-thought reasoning combined with few-shot learning to enhance the analysis and reasoning capabilities for understanding the semantics of functions.

\textbf{(1) LLM-based APR.} 
Unlike the specifically designed NMT-based APR models, LLMs are unsupervised trained using up to billions of text/code tokens.
This large-scale unsupervised learning process allows LLMs to have strong reasoning capabilities and be applied for program repair without relying on training with a large amount of historical bug fixes.
Therefore, we propose a novel self-directed LLM-based APR, namely \toolname, since representative conversational LLM model provides advanced capabilities for several tasks, including bug-fixing~\cite{xia2023keep, sobania2023analysis, haque2023potential}.

\textbf{(2) Automated few-shot CoT for APR.}
Automated program repair is not a trivial task since it requires logical thinking to understand the semantics of the buggy function and a coherent series of intermediate steps to fix the bugs. 
Even with powerful reasoning capabilities, LLMs still require some guidance in orchestrating the steps for fixing. 
Instead of directly generating fixed code, chain-of-thought inside LLMs can help analyze and reason the code logic.
Meanwhile, few-shot examples can help LLMs to better under the faced task, but the quality of examples will highly affect the capabilities. 
Thus, we design an automated approach that extracts the chains of thoughts from the LLMs, selects effective examples for few-shot learning, and composes a prompt with CoTs for fixing. 
\label{sec:motivate}

\vspace{-0.2cm}
\section{Our Approach: \toolname}

\xy
{
{\bf \toolname} has two main phases (illustrated in Fig.~\ref{fig:overview} and Algorithm~\ref{alg}): \ding{182} \textbf{collection phase} and \ding{183} \textbf{fixing phase}.
The former phase is to collect chains of thoughts that constitute pre-fixed knowledge, and the latter phase fixes a bug with CoT-based prompting and few-shot learning. 
In this paper, we adopt ChatGPT~\cite{openai2022chatgpt}, CodeLlama~\cite{roziere2023code}, DeepSeek-Coder~\cite{deepseek-coder}, and StarCoder~\cite{li2023starcoder} as the backend LLMs. 
\toolname is flexible to include other state-of-the-art LLMs as the backend model.
The details of \toolname are presented in the following subsections.
}

\IncMargin{-0.5em}
\begin{algorithm} 
\small
    \textbf{Collection Phase (Section \ref{sec:collection})}
    
    \textbf{Input: }Buggy functions ${F}$;
    
    \For{$f$ in $F$} {
        \emph{1: Combine the buggy function $f$ into the prompt;}
        
        \emph{2: Generate CoT and fixed function by prompting the LLM;}
        
        \emph{3: Test the fixed function, retaining the valid function (with their buggy function and CoT) into ${K}$;}
    }
    
    {\textbf{Output: }Knowledge Pool ${K}$;}
    
    \vspace{0.1cm} 
    
    \hrule

    \vspace{0.1cm} 
    
    \textbf{Fixing Phase (Section \ref{sec:fixing})}
    
    \textbf{Input: }Knowledge Pool ${K}$, Buggy function $f$, interaction=1;
        
    \emph{1: Select few-shot examples $E$ from knowledge pool ${K}$;}

    \emph{2: Combine the example $E$ and buggy function $f$ into the prompt;}
    
    \emph{3: Generate CoT and fixed function by prompting the LLM;}

    \emph{4: \While{Fixed Function is invalid \&\& interaction++ < 5} {
        Add test failure information to prompt and regenerate;
    }}
    
    \textbf{Output: }Fixed function $f'$;
    \caption{\xy{Collection Phase and Fixing Phase}}
    \label{alg}
\end{algorithm}

\vspace{-0.2cm}
\subsection{Collection Phase}
\label{sec:collection}

This phase aims to collect a variety of chains of thoughts that constitute the knowledge pool.
To achieve this, we need to address three tasks: \textbf{(1) Prompt Preparation}, \textbf{(2) Chains of Thoughts Collection}, and \textbf{(3) Function Verification}.

\subsubsection{Task 1: Prompt Preparation} 
\begin{figure}[htbp]
    \centering
    \vspace{-0.2cm}
    \includegraphics[width=.70\linewidth]{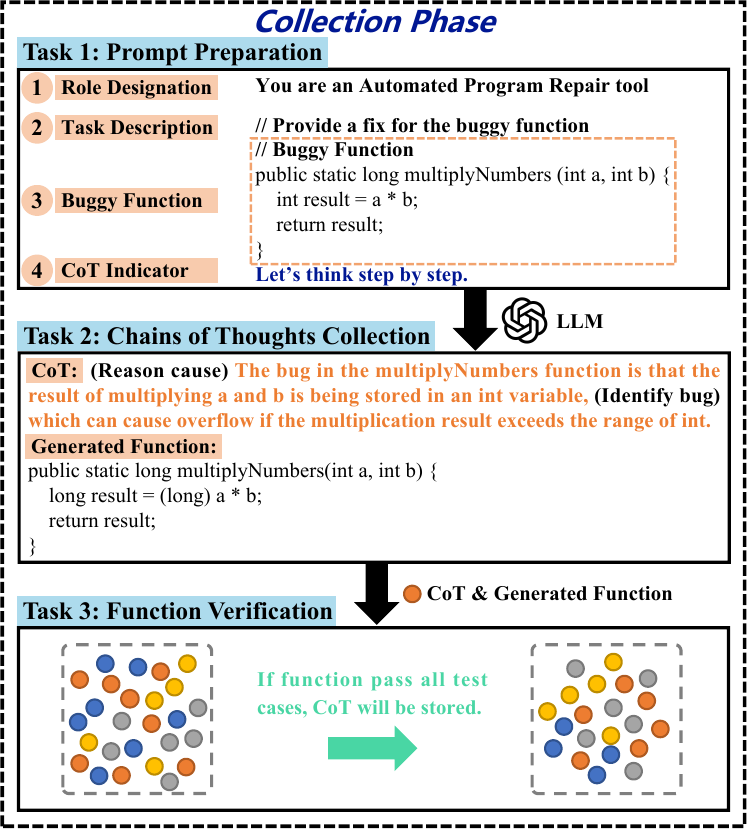}
    \caption{The Process of the Collection Phase} 
    \label{fig:collect}
    \vspace{-0.5cm}
\end{figure}

The prompt used in \toolname involves four important components as illustrated in Fig.~\ref{fig:collect}:

\begin{itemize}[leftmargin=*]
\item \textbf{Role Designation} (marked as \ding{172}). 
\toolname starts a role for LLM with an instruction like \texttt{``You are an Automated Program Repair tool''}.

\item \textbf{Task Description} (marked as \ding{173}). 
LLM is provided with the description constructed as \texttt{``// Provide a fix for the buggy function''}. 
Since we illustrate an example in Java, we use the Java comment format of \texttt{``//''} as a prefix.

\item \textbf{Buggy Function} (marked as \ding{174}). 
\toolname provides the buggy function 
to LLM 
in our single-function fixing scenario. 
We also prefix the buggy function with \texttt{``// Buggy Function''} to directly indicate LLM about the context of the function.

\item \textbf{Chain-of-Thought Indicator} (marked as \ding{175}).
LLM is instructed to think step-by-step when fixing a bug.
In this paper, we follow the best practice in previous work~\cite{wei2022chain} and adopt the same prompt named \texttt{``Let's think step by step''}.
\end{itemize}

\subsubsection{Task 2: Chains of Thoughts Collection}
Given a corpus of buggy functions, \toolname uses the decorated prompt to collect chains of thoughts on fixing the buggy functions. 
The output of Task 2 is a collection of samples, where a sample includes a buggy function, its fixed version, and the chain of thought.
As an example in Fig.~\ref{fig:collect}, the \textit{multiplyNumbers} function initially fails to consider the possibility of an overflow bug when performing multiplication operations.
LLM reasons that the result of multiplying \textit{``integer a''} and \textit{``integer b''} inside \textit{``multiplyNumbers''} function can cause a bug. 
The result is originally planned to be stored in another integer variable, which may cause an overflow bug if the multiplication result exceeds the range of an integer. 
Thus, it suggests converting the type of multiplication results (i.e., \textit{``result''}) into a \textit{``long''} type and consequently resolving the overflow problem.

\subsubsection{Task 3: Function Verification}
To get effective samples, it is imperative to filter out low-quality thought processes.
\toolname runs a test suite (originally supported by the studied dataset, cf. Section~\ref{sec:dataset}) to test the fixed functions extracted from LLM's output in Task 2, retaining only the fixed functions (with their buggy functions and CoTs) that successfully pass the entire test suite.
Meanwhile, LLM may not always correctly fix one buggy function at the first attempt. 
Thus, we execute the process at most 25 attempts for each bug (refer to Section~\ref{sec:setting} for more details).

\subsection{Fixing Phase}
\label{sec:fixing}

In this phase, \toolname first selects diverse and effective samples from the knowledge pool in the collection phase.
\toolname automatically utilizes selected examples (i.e., buggy function as well as its corresponding fixed version appended with reasoning process) and the targeted buggy function (i.e., function to be fixed) to compose a prompt to interact with LLM. 
Finally, \toolname obtains the output from LLM including both the chains of thoughts and the candidate fixed function to the buggy one.
Notice that each candidate function generated by LLM will be passed through a function verification step and the feedback from the verification step will be appended to LLM for further refinement.
Overall, the fixing phase has three tasks: \textbf{(1) Few-Shot Selection}, \textbf{(2) Automatic Fixing}, and \textbf{(3) Interaction Verification}.

\begin{figure}[htbp]
    \centering
     \vspace{-0.2cm}
    \includegraphics[width=.84\linewidth]{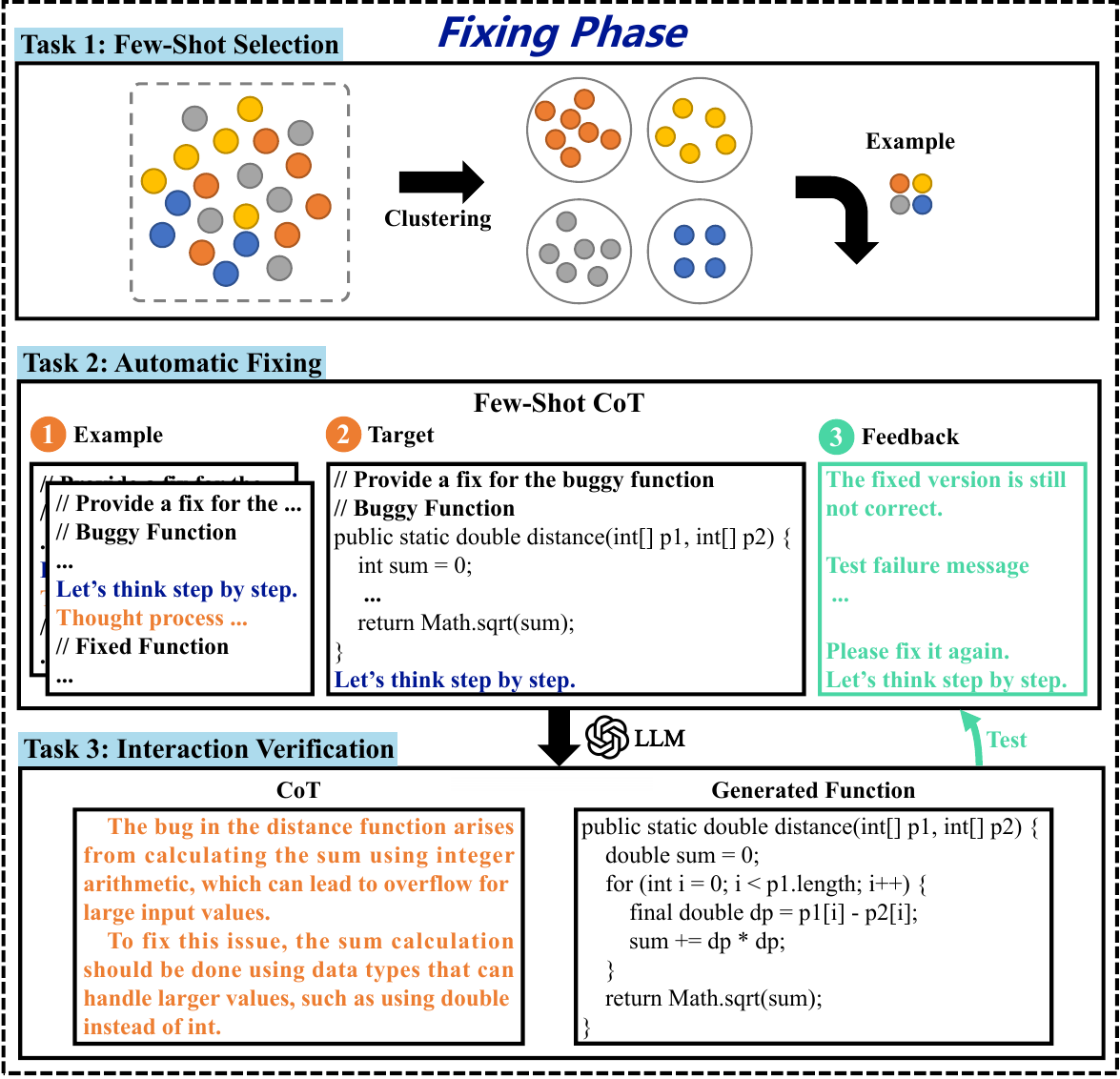}
    \caption{
    The Process of the Fixing Phase}
    \label{fig:inference}
     \vspace{-0.5cm}
\end{figure}

\subsubsection{Task 1: Few-Shot Selection}
\label{sec:select}

LLM needs a high-quality prompt to instruct itself to finish the downstream tasks, which is also the focus of prior works~\cite{xia2023keep,white2023prompt,feng2023prompting}. 
Similarly, we aim to reduce labor involvement in instructing LLM by guiding it to learn from the solved problems.
To achieve this, we need to identify the most beneficial examples from the knowledge pool. 
Moreover, previous work~\cite{zhang2022automatic} concludes that a few diverse examples may help LLM achieve a better generalization ability.
Therefore, \toolname clusters these examples inside the knowledge pool on the basis of their semantic similarity to pick out distinct ones~\cite{zhang2022automatic, li2023mot}.
Furthermore, considering the limitations of LLMs' conversation windows, all examples in the knowledge pool are clustered into two clusters, with one sample selected from each cluster (i.e., two shots used in this paper).
\xy
{
We employ two advanced embedding strategies (i.e., Semantic-based Selection and Contrastive-based Selection) to select semantically similar examples.
For comparison, we also use IR-based Selection and Randomly Selection, and the details are elaborated as follows.
}

\begin{itemize}[leftmargin=*]
\item \xy{\textbf{Semantic-based Selection} adopts a pre-trained model (i.e., UniXcoder, which effectively comprehends code semantic information~\cite{guo2022unixcoder, ni2023fva}) to embed all the buggy functions and then uses the K-means algorithm~\cite{macqueen1967some} for clustering. 
During the fixing phase, the most semantic similar examples are picked out from each cluster based on cosine similarity.}

\item \xy{\textbf{Contrastive-based Selection} utilizes the contrastive learning framework R-Drop~\cite{ni2023distinguishing} to further fine-tune UniXcoder for better semantic embedding.
We input one function twice to get the embeddings $E_1$ and $E_2$.
The training objective is the distance between $E_1$ and $E_2$ should be as small as possible.
For clustering, it remains the same operation as Semantic-based Selection.}

\item \xy{\textbf{IR-based Selection} builds indexes for code in the knowledge pool and then retrieves similar examples by BM25 score~\cite{robertson2009probabilistic}.}

\item \textbf{Randomly Selection} randomly selects a few examples from the knowledge pool built in the collection phase.

\end{itemize}

\subsubsection{Task 2: Automatic Fixing}

\toolname utilizes the selected examples and the target buggy function (i.e., to be fixed) to construct a prompt (i.e., \ding{172} + \ding{173} marked in Fig.~\ref{fig:inference}).
Then, \toolname uses this prompt to interact with LLM and help it to infer the bug-fixing solution.
Following that, we can obtain the model outputs, which usually contain the process of LLM's thought and the candidate fixed function, which will be verified in the next step.

\subsubsection{Task 3: Interaction Verification}
\toolname complies and runs test suite to verify all candidate fixed functions generated by LLM. 
In case a candidate function fails to pass all test cases, \toolname first collects the failing test information, which can aid LLM in understanding the failure causes and provide guidance to generate the correct fix.
In particular, the test failure messages can be divided into four categories: 
(1) \textit{Compile Fail}, (2) \textit{Time Out}, (3) \textit{Syntax Error}, and (4) \textit{Failing test}: \textit{TestClass::TestFunction}.
Then, \toolname reconstructs the prompt and appends the failing information (i.e., \ding{174} marked in Fig.~\ref{fig:inference}) to the back of the original prompt (i.e., \ding{172} + \ding{173} marked in Fig.~\ref{fig:inference}).
Here, the failing information is added into a template: \textit{``The fixed version is still not correct. \{test failure message\}. Please fix it again. Let's think step by step.''}
Then, \toolname interacts with LLM using the new prompt (i.e., \ding{172}+\ding{173}+\ding{174}) to generate a new fixed solution.
LLM can avoid generating similar mistakes and also learn from previous interactions based on new prompt.
This iterative process continues until a fixed function is obtained (i.e., successfully passing the entire test suite) or the maximum number of interactions exceeds (i.e., five times for a better balance between effectiveness and cost, refer to Section~\ref{sec:rq3}).
\label{sec:method}

\section{Experimental Design}

We present the experimental design, including studied datasets, baselines, evaluation metrics, and experiment settings.

\subsection{Datasets}
\label{sec:dataset}

For the evaluation, we follow previous studies~\cite{xia2023automated, xia2022less, xia2023keep} to adopt the two APR benchmarks (\textbf{Defects4J dataset}~\cite{just2014defects4j} and \textbf{QuixBugs dataset}~\cite{lin2017quixbugs}) spanning across two popular programming languages (i.e., Java and Python). 
Similar to prior APR studies ~\cite{xia2023automated, xia2023keep, xia2022less}, we separate Defects4J into Defects4J 1.2 and Defects4J 2.0. 
Defects4J 1.2 consists of 391 bugs in 6 different Java projects, while Defects4J 2.0 consists of 438 new bugs across 9 additional projects.
We also focus on scenarios where the fix is solely located in a single function since it is the focal point of most recent APR work ~\cite{xia2022less, xia2023automated, ye2022neural, zhu2021syntax, jiang2021cure, lutellier2020coconut}. 
We filter out the datasets to contain single function bugs.
The statistic of each evaluation dataset is presented in Table ~\ref{tab:dataset}.

\begin{table}[!htbp]
    \vspace{-0.1cm}
    \centering
    \caption{\xy{Statistics of studied dataset}}
    \resizebox{\linewidth}{!}
    {
        \begin{threeparttable}
        \begin{tabular}{lrrrr}
        \toprule
        \textbf{Dataset} & \textbf{\# Total Bugs} & \textbf{\# SF Bugs} & \textbf{\xy{\# SH Bugs}} & \textbf{\xy{\# SL Bugs}} \\
        \midrule
        Defects4J 1.2 & 391 & 255 & \xy{154} & \xy{80}  \\
        Defects4J 2.0 & 438 & 228 & \xy{159} & \xy{78} \\
        QuixBugs-Java & 40 & 40 & \xy{37} & \xy{37} \\
        QuixBugs-Python & 40 & 40 & \xy{40} & \xy{40} \\
        \midrule
        \textbf{\# Sum} & \textbf{909} & \textbf{563} & \xy{\textbf{390}} & \xy{\textbf{235}}\\
        \bottomrule
        \end{tabular}
        $^\ast$Defects4J 1.2 and Defects4J 2.0 are two completely independent versions, with no overlapping bugs between them.
        \end{threeparttable}
    } 
    \vspace{-0.3cm}
    \label{tab:dataset}
\end{table}

\subsection{Baselines and Evaluation Metrics}

{\bf Studied Baselines.} 
We compare \toolname with the twelve state-of-the-art APR approaches in Table~\ref{tab:baselines}, including eight NMT-based and four LLM-based SOTAs in APR. 
AlphaRepair is an LLM-based repair tool that employs the pre-trained CodeBERT model ~\cite{fengetal2020codebert} with cloze-style APR, which means that it does not require fine-tuning on bug-fixing data.
Codex and GPT-NeoX directly applied LLMs for APR without fine-tuning~\cite{xia2023automated}.
ChatRepair is similar to our work, which repairs a bug with a conversational ChatGPT. 
The NMT-based APR models require fine-tuning for better performance, while LLM-based APR approaches do not require fine-tuning.



\begin{table}[htbp]
    \centering
     \vspace{-0.3cm}
    \caption{The studied baselines}
    \resizebox{\linewidth}{!}{
    \begin{threeparttable}     
        \begin{tabular}{lccrr}
        \toprule
        \textbf{Approach} & \textbf{Category} & \textbf{Fine-Tuning} & \textbf{Time/\# Patch} & \textbf{Venue} \\
        \midrule
        KNOD~\cite{jiang2023knod} & NMT-based & \ding{51} & 5 Hour & ICSE 2023 \\
        TENURE~\cite{meng2023tenure} & NMT-based & \ding{51} & 5 Hour & ICSE 2023 \\
        RewardRepair~\cite{ye2022neural} & NMT-based & \ding{51} & 200 & ICSE 2022 \\
        SelfAPR~\cite{ye2022selfapr} & NMT-based & \ding{51} & no setting reported & ASE 2022 \\
        CURE~\cite{jiang2021cure} & NMT-based & \ding{51} & 10,000 & ICSE 2021 \\
        DeepDebug~\cite{drain2021deepdebug} & NMT-based & \ding{51} & 100 & arXiv 2021 \\
        CoCoNuT~\cite{lutellier2020coconut} & NMT-based & \ding{51} & 10,000 & ISSTA 2020 \\
        DLFix~\cite{li2020dlfix} & NMT-based & \ding{51} & 5 Hour & ICSE 2020 \\
        \midrule
        ChatRepair~\cite{xia2023keep} & LLM-based & \ding{55} & $\leq 200$ & arXiv 2023 \\
        Codex~\cite{xia2023automated} & LLM-based & \ding{55} & 200 & ICSE 2023 \\
        GPT-NeoX~\cite{xia2023automated} & LLM-based & \ding{55} & 200 & ICSE 2023 \\
        AlphaRepair~\cite{xia2022less} & LLM-based & \ding{55} & 5,000 & FSE 2022 \\
        \rowcolor{lightgray} \textbf{Ours: \toolname} & LLM-based & \ding{55} & $\leq 125$ & This work \\
        \bottomrule
        \end{tabular}%
        ``\textbf{Time/\# Patch}'': indicates the longest time set for fixing a bug or the maximum number of candidate patches that a model was set to loop through via running test cases before a correct patch is obtained.
    \end{threeparttable}
    } 
    \vspace{-0.3cm}
    \label{tab:baselines}%
\end{table}%

\xy
{
{\bf Evaluation Metrics.} Following previous work~\cite{jiang2023knod, meng2023tenure, xia2023automated, xia2022less, ye2022selfapr, jiang2021cure, ye2022neural, lutellier2020coconut, li2020dlfix,li2022dear,li2020improving}, we adopt two widely used metrics for evaluating approaches: \textit{(1) number of correct patches} and \textit{(2) number of plausible patches}. 
A plausible patch is a patch that can pass all test cases but is not semantically equivalent to the actual fix.
Following the previous work~\cite{xia2023automated, xia2022less}, we also manually check and identify the plausible patches that are semantically equivalent to the actual fixes.
}


\begin{table*}[htbp]
  \centering
  \vspace{-0.2cm}
  \caption{RQ1: \toolname vs. Basic LLMs for different projects on Defects4J V1.2}
  \resizebox{.73\linewidth}{!}
  {
    \begin{threeparttable}
    \begin{tabular}{l|cccc|cccc}
    \toprule
    \multicolumn{1}{l|}{\multirow{2}[2]{*}{\textbf{Projects}}} & \multicolumn{4}{c|}{\textbf{\toolname with perfect fault info}} & \multicolumn{4}{c}{\textbf{Basic LLM with perfect fault info}} \\
\cmidrule{2-9}    \multicolumn{1}{l|}{} & \textbf{ChatGPT} & \textbf{CodeLlama} & \textbf{DeepSeek} & \textbf{StarCoder} & \textbf{BaseChatGPT} & \textbf{BaseCodeLlama} & \textbf{BaseDeepSeek} & \textbf{BaseStarCoder} \\
    \midrule
    \textbf{Chart} & \cellcolor{lightgray}11 & 10 & 10 & 8 & 5 & 6 & 6 & 5 \\
    \textbf{Closure} & \cellcolor{lightgray}31 & 20 & 14 & 20 & 13 & 11 & 9 & 14 \\
    \textbf{Lang} & \cellcolor{lightgray}19 & 13 & 11 & 12 & 11 & 10 & 7 & 7 \\
    \textbf{Math} & \cellcolor{lightgray}27 & 21 & 20 & 16 & 16 & 12 & 13 & 9 \\
    \textbf{Mockito} & \cellcolor{lightgray}6 & 4 & 4 & 2 & \cellcolor{lightgray}6 & 4 & 3 & 2 \\
    \textbf{Time} & \cellcolor{lightgray}4 & 2 & 4 & 1 & 1 & 1 & 1 & 0 \\
    \midrule
    \textbf{\# Sum} & \cellcolor{lightgray}98 & 70 & 63 & 59 & 52 & 44 & 39 & 37 \\
    \bottomrule
    \end{tabular}%
    \end{threeparttable}
    }
    \vspace{-0.3cm}
  \label{tab:vs_base_llm}%
\end{table*}%

\vspace{-0.1cm}
\subsection{Implementation}
\label{sec:setting}

We developed the generation pipeline in Python, utilizing PyTorch~\cite{pytorch} implementations of CodeLlama 13B, DeepSeek-Coder 7B, and StarCoder 16B. 
We use the Hugging Face~\cite{huggingface} to load the model weights and generate outputs.
For ChatGPT, we utilize OpenAI's API access~\cite{2023chatgptendpoint}, complying with the recommended best practices~\cite{shieh2023best} for each prompt. 
We utilize the gpt-3.5-turbo model from the ChatGPT family, which is the version used uniformly for our experiments.
A sampling temperature of 1 is utilized to obtain a diverse set of potential patches~\cite{gilardi2023chatgpt}.
The maximum number of repair attempts is set to 25 (i.e., 25 independent sessions) for \toolname as the default for auto-fixing the single function bugs.
Since LLM has a maximum input limit, we employ two few-shot examples and set the maximum interaction number to 5 (i.e., up to 5 interactions in a session).
The evaluation is conducted on a 16-core workstation equipped with an Intel(R) Xeon(R) Gold 6226R CPU @ 2.90Ghz, 192GB RAM and NVIDIA RTX 3090 GPU, running Ubuntu 20.04.1 LTS.
Following previous APR works~\cite{xia2022less, zhu2021syntax}, we use a default end-to-end timeout of 5 hours to fix one bug. 
In practice, the total time required is on average lower than 20 minutes since we sample small-scale patches (i.e., 25 attempts $\times$ 5 interactions) for each bug.
\label{sec:experiment}

\vspace{-0.3cm}
\section{Experimental Results}

To investigate the effectiveness of \toolname on bug fixing, our experiments focus on the following three research questions:

\begin{itemize}[leftmargin=*]
\item \textbf{RQ-1 Comparable Study on LLM-based APRs.} {\em How does the performance of \toolname compare with the LLM-based APRs?}

\item \textbf{RQ-2 Comparable Study on NMT-based APRs.} {\em How does the performance of \toolname compare with the state-of-the-art NMT-based APRs?}

\item \textbf{RQ-3 Sensitivity Analysis.} {\em How do different configurations affect the overall performance of \toolname?}
\end{itemize}

\vspace{-0.3cm}
\subsection{RQ-1: Compare with LLM-based APRs}
\label{sec:rq1}

\noindent
\textbf{\underline{RQ1-Analysis Procedure.}}
We evaluate \toolname against four LLM-based APRs: ChatRepair~\cite{xia2023keep}, Codex~\cite{xia2023automated}, GPT-NeoX~\cite{xia2023automated}, and AlphaRepair~\cite{xia2022less}. 
We adopt ChatGPT~\cite{openai2022chatgpt}, CodeLlama~\cite{roziere2023code}, DeepSeek-Coder~\cite{deepseek-coder} (we refer to as DeepSeek), and StarCoder~\cite{li2023starcoder} as the backend LLMs for \toolname.
In addition, we build baseline approaches, BaseChatGPT, BaseCodeLlama, BaseDeepSeek, and BaseStarCoder, which utilize the basic LLMs to perform a repair (i.e., directly using existing bug-fixing data as examples) without any Chains-of-Thoughts reasoning process and feedback information.

{\bf Data Splitting.} Our \toolname has two phases: the \textit{Knowledge Collection} and \textit{Fixing}. 
We use the buggy functions in Defects4J V2.0 (i.e., 228) for collection and the ones in Defects4J V1.2 (i.e., 255) for the fixing phase. 
Defects4J V1.2 and Defects4J V2.0 are two completely independent versions, with no overlapping bugs between them.
In addition, we also conduct an experiment to use functions in Defects4J V1.2 for collection and the ones in Defects4J V2.0 for the fixing phase.
Since QuixBugs dataset has the limited number of functions (i.e., 40 for both Java and Python), for comprehensively evaluating the performance differences, we adhere to the best-practice guide~\cite{shieh2023best}, manually designed 2 examples for fixing bugs without the collection phase.

{\bf Fault Information.} Following previous work~\cite{xia2022less, xia2023automated, ye2022neural, jiang2021cure, lutellier2020coconut}, we focus on two settings: (1) No fault localization is performed, the perfect fault information (i.e., including statement-level fault information) is provided; (2) The fault information at method-level is provided but no statement-level fault information.

{\bf Experiment Settings.} Since all of the studied baselines were already evaluated on Defects4J, following the convention in APR work~\cite{xia2022less, xia2023automated}, we directly compare the auto-fix results obtained in previous studies~\cite{xia2023automated, xia2023keep, xia2022less} under \underline{the same settings.}

As for the few-shot selection strategy during the fixing phase, we adopt the \textit{Contrastive-based Selection} strategy since it has the overall best performance. 
Meanwhile, we set the number of interaction feedback as five since it achieves a better balance between effectiveness and cost (cf. Section~\ref{sec:rq3}).

In \toolname, we study 3 different repair scenarios used in previous works~\cite{xia2023automated, xia2023keep}: \textit{single function}, \textit{single hunk}, and \textit{single line}.
Note that \textit{single hunk} fix is a subset of \textit{single function} fix and \textit{single line} fix is a subset of \textit{single hunk} fix.
In QuixBugs-Java, \textit{single hunk} is equal to \textit{single line} (i.e., all \textit{single hunk} bugs require fixing just one line of code), while in QuixBugs-Python, all fixes are \textit{single line}.

The experimental settings for \toolname: only \textit{single function} scenario is considered, the results for \textit{single hunk} and \textit{single line} come from \textit{single function} results. 
Whereas, the experimental settings for ChatRepair, Codex, and GPT-NeoX are that \textit{single function}, \textit{single hunk}, and \textit{single line} scenarios are experimented separately, and the results of three experiments are independent to each other. 
These models conducted experiments on the single function bugs in the same setting as ours, we therefore compared with them on single function setting only.

\begin{table}[htbp]
    \centering
    \caption{RQ1: \toolname vs. LLM-based APRs (SF: Single Function, SH: Single Hunk, SL: Single Line)}
    \resizebox{.8\linewidth}{!}{
    \begin{threeparttable}
        \begin{tabular}{lccc|ccc|cc|c}
        \toprule
        \multirow{3}[5]{*}{\textbf{Models}} & \multicolumn{6}{c}{\textbf{Defects4J}} & \multicolumn{3}{c}{\textbf{QuixBugs}} \\
        \cmidrule{2-10} & \multicolumn{3}{c}{\textbf{V1.2}} & \multicolumn{3}{c|}{\textbf{V2.0}} & \multicolumn{2}{c}{\textbf{Java}} & \multicolumn{1}{c}{\textbf{Python}} \\
        \cmidrule{2-10} & \textbf{SF} & \textbf{SH} & \textbf{SL} & \textbf{SF} & \textbf{SH} & \textbf{SL} & \textbf{SF} & \textbf{SH} & \textbf{SF} \\
        \midrule
        \multicolumn{10}{l}{\textbf{\footnotesize Method-level fault info.}} \\
        Codex & 63 & - & - & - & - & - & 32 & - & 37 \\       
        GPT-NeoX & 18 & - & - & - & - & - & 8 & - & 19  \\  
        BaseChatGPT & 36 & 28 & 20 & 39 & 29 & 23 & 38 & 36& 35 \\
        ThinkRepair* & {63} & {46} & {33} & {62} & {44} & {25} & {38} & {36} & {38} \\
        \rowcolor{lightgray} \toolname & {80} & {64} & {44} & {90} & {69} & {41} & {39} & {36} & {40} \\
        \midrule
        \multicolumn{10}{l}{\textbf{\footnotesize Perfect fault info.}} \\
        AlphaRepair & 67 & 57 & 48 & - & - & 35 & 28 & 27 & 27 \\
        ChatRepair & 76 & - & - & - & - & - & {39} & - & {40} \\ 
        BaseChatGPT & 52 & 44 & 31 & 46 & 35 & 25 & 38 & 36 & 37 \\
        ThinkRepair* & {70} & {55} & {37} & {72} & {53} & {36} & {38} & {36} & {38} \\
        \rowcolor{lightgray} \toolname & {98} & {78} & {52} & {107} & {81} & {47} & {39} & {36} & {40} \\
        \bottomrule
        \end{tabular}%
        ``-'': indicates no results reported in the original work, or cannot be directly compared since different experimental settings.
    \end{threeparttable}
    } 
    \vspace{-0.5cm}
    \label{tab:vs_llm}
\end{table}

\begin{table*}[htbp]
  \centering
  \vspace{-0.2cm}
  \caption{RQ1: \toolname vs. LLM-based APRs for different projects on Defects4J V1.2}
  \resizebox{.83\linewidth}{!}
  {
    \begin{threeparttable}
    \begin{tabular}{l|ccccc|ccccc}
    \toprule
    \multicolumn{1}{l|}{\multirow{2}[2]{*}{\textbf{Projects}}} & \multicolumn{5}{c|}{\textbf{Method-level fault info}} & \multicolumn{5}{c}{\textbf{Perfect fault info}} \\
    \cmidrule{2-11} 
    \multicolumn{1}{l|}{} & \textbf{ThinkRepair} & \textbf{ThinkRepair*} & \textbf{BaseChatGPT} & \textbf{GPT-NeoX} & \textbf{Codex} & \textbf{ThinkRepair} & \textbf{ThinkRepair*} & \textbf{BaseChatGPT} & \textbf{ChatRepair} & \textbf{AlphaRepair} \\
    \midrule
    \textbf{Chart} & \cellcolor{lightgray}9 & 8 & 3 & - & - & \cellcolor{lightgray}11 & 10 & 5 & - & 8 \\
    \textbf{Closure} & \cellcolor{lightgray}19 & 19 & 5 & - & - & \cellcolor{lightgray}31 & 20 & 13 & - & 22 \\
    \textbf{Lang} & \cellcolor{lightgray}15 & 11 & 7 & - & - & \cellcolor{lightgray}19 & 13 & 11 & - & 11 \\
    \textbf{Math} & \cellcolor{lightgray}27 & 20 & 16 & - & - & \cellcolor{lightgray}27 & 21 & 16 & - & 19 \\
    \textbf{Mockito} & \cellcolor{lightgray}7 & 4 & 4 & - & - & \cellcolor{lightgray}6 & 4 & 6 & - & 4 \\
    \textbf{Time} & \cellcolor{lightgray}3 & 1 & 1 & - & - & \cellcolor{lightgray}4 & 2 & 1 & - & 3 \\
    \midrule
    \textbf{\# Sum} & \cellcolor{lightgray}80 & 63 & 36 & 18 & 63 & \cellcolor{lightgray}98 & 70 & 52 & 76 & 67 \\
    \bottomrule
    \end{tabular}%
    ``-'': indicates no results reported in the original work~\cite{xia2023automated, xia2023keep}.
    \end{threeparttable}
    } 
    \vspace{-0.3cm}
  \label{tab:vs_llm_project}%
\end{table*}%

\noindent
\textbf{\underline{RQ1-Results.}} 
\textbf{\toolname vs. Basic LLMs.} 
Table~\ref{tab:vs_base_llm} illustrates the number of bugs successfully repaired by \toolname and Basic LLMs for different projects in the scenario of \textit{single function} bug-fixing. 
We observe that with perfect fault information, \toolname demonstrates superior performance across all projects compared to Basic LLMs.
In particular, the performance of \toolname is significantly better than Basic LLM with improvements ranging from 59.1\%$\sim$88.5\%. 
This not only demonstrates the superiority of our \toolname approach, but also its universal applicability, as it is not tailored for any specific LLM, but is suitable for various LLMs. 
This underscores its LLM-agnostic design paradigm.
To facilitate comparison, we employ the top two performing models, ChatGPT and CodeLlama, in the following sections of our results, denoting them as \underline{\toolname and ThinkRepair*}, respectively.

\textbf{\toolname vs. ChatGPT-based APRs.}
With the perfect fault information provided, \toolname can auto-fix 98 bugs and 22 more bugs (28.9\% improvements) than ChatRepair for single function bugs in Defects4J V1.2.
For the QuixBugs dataset, \toolname and ChatRepair can auto-fix the same number of bugs. 
Both models can auto-fix all bugs in QuixBugs-Python, but miss one bug in QuixBugs-Java.

\toolname can outperform the BaseChatGPT in all settings.
For example, with the method-level fault information, \toolname can auto-fix 122.2\%, 128.6\%, and 120\% more bugs than BaseChatGPT for single function, single hunk, and single line bugs in Defects4J V1.2, respectively. 
When provided with perfect fault information, \toolname can auto-fix 88.5\%, 77.3\%, and 67.7\% more bugs for single function, single hunk, and single line bugs in Defects4J V1.2.
Similarly, on Defects4J V2.0, \toolname achieves an overwhelming better performance than BaseChatGPT.

Table~\ref{tab:vs_llm_project} illustrates the number of bugs successfully repaired by \toolname and LLM-based APRs for different projects in the scenario of \textit{single function} bug-fixing. 
\toolname and ThinkRepair* represent for the results of our approach with ChatGPT and CodeLlama as the backend LLMs, respectively.
We observe that with or without perfect fault information, \toolname demonstrates superior performance across all projects compared to BaseChatGPT.
For example, for the project Closure, \toolname improves the BaseChatGPT by 280\% and 138.5\% with the method-level and perfect fault information, respectively.

{\bf \toolname vs. other LLM-based APRs.}
For single function bugs in Defects4J V1.2, \toolname can auto-fix 31 more bugs (46.3\%) than AlphaRepair with the perfect fault information provided, and auto-fix 17 (27\%) and 62 (344.4\%) more bugs than Codex and GPT-NeoX with only method-level fault information provided, respectively. 
As for QuixBugs, \toolname obtains the best result with the perfect and method-level fault information. 
As shown in Table~\ref{tab:vs_base_llm} and Table~\ref{tab:vs_llm}, BaseCodeLlama demonstrates relatively poor bug-fixing capabilities.
However, under our unsupervised setting, CodeLlama (i.e., ThinkRepair*) can achieve performance on par with or surpass supervised approach (i.e., AlphaRepair).

By observing Table~\ref{tab:vs_llm_project}, we find that with perfect fault information, \toolname outperforms AlphaRepair in all projects.
In particular, the performance of \toolname is significantly better than AlphaRepair in three projects (Closure, Lang, and Math), with improvements ranging from 8 to 9 bugs (40.9\%$\sim$72.7\%).

\begin{figure}[htbp]
    \centering
     \vspace{-0.2cm}
    \includegraphics[width=.7\linewidth]{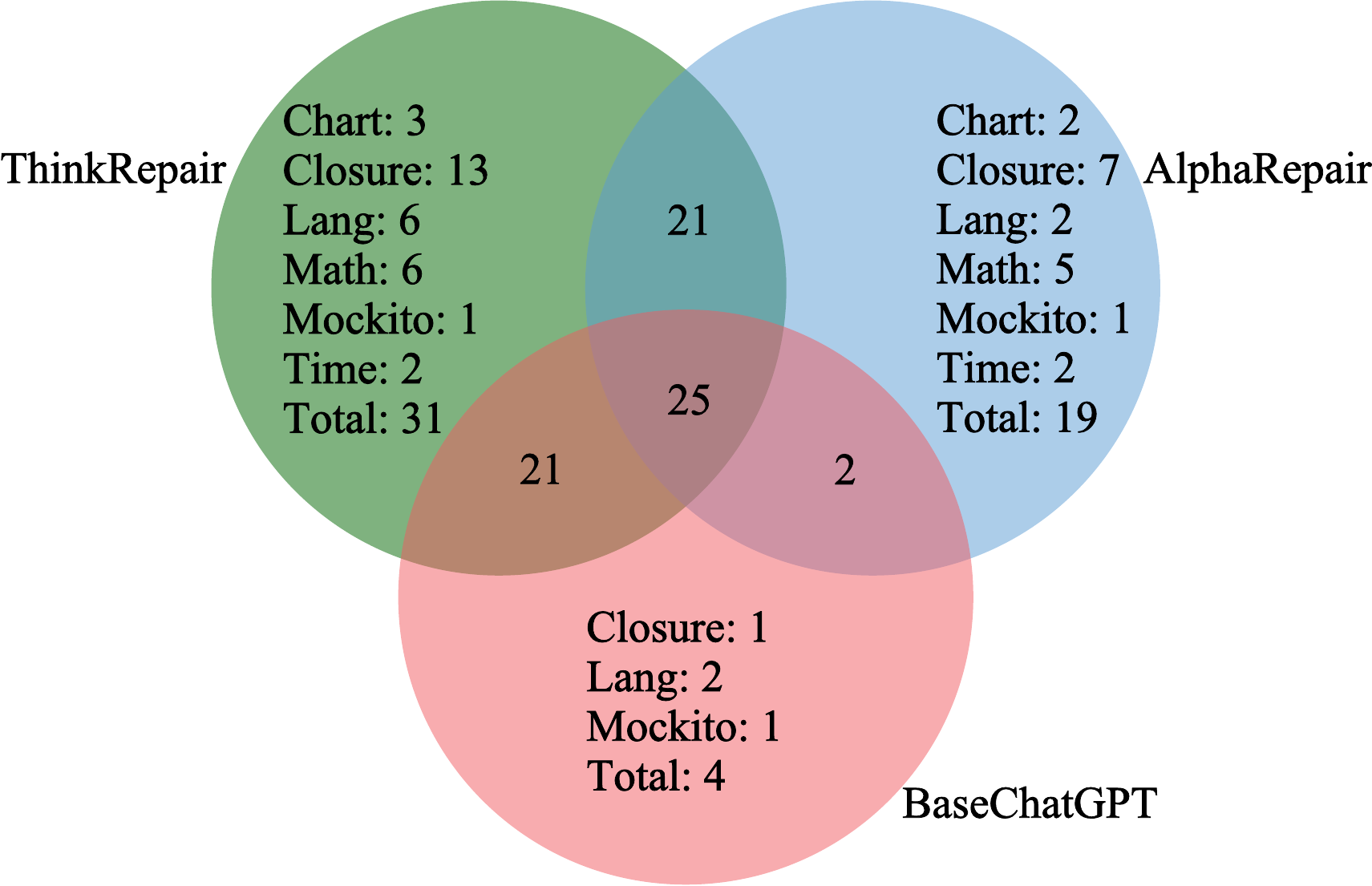}
    \caption{RQ1: Bug-fixing Venn diagram on Defects4J V1.2 of \toolname, BaseChatGPT and AlphaRepair}
     \vspace{-0.2cm}
    \label{fig:venn_llm}
\end{figure}

Fig.~\ref{fig:venn_llm} illustrates the Venn diagram depicting the bugs fixed by \toolname, BaseChatGPT, and AlphaRepair on Defects4J V1.2.
It is noteworthy that \toolname successfully fixes 31 unique bugs that BaseChatGPT and AlphaRepair are unable to resolve.
Meanwhile, BaseChatGPT can exclusively correctly fix four bugs (i.e., \href{https://storage.googleapis.com/google-code-archive/v2/code.google.com/closure-compiler/issues/issue-501.json}{Closure-61}, \href{https://issues.apache.org/jira/browse/LANG-346}{Lang-53}, \href{https://issues.apache.org/jira/browse/LANG-304}{Lang-57} and \href{https://github.com/mockito/mockito/issues/188}{Mockito-12}), which \toolname fails to fix.
The reason may be attributed to \toolname's inclination to excessively modify bugs stemming from over-inferring.

\xy
{
A specific example is \href{https://github.com/mockito/mockito/issues/188}{Mockito-12}, where \toolname incorrectly revises the non-buggy line \textit{``if (generic != null \&\& generic instanceof ParameterizedType)''} to \textit{``if (generic instanceof ParameterizedType''} even though it has passed all test suites after adding the \textit{``if (num.intValue() < 0)''} and \textit{``else if (actual instanceof ParameterizedType)''} condition.
As a consequence, this modification results in divergent semantics compared to the correct code repair.
}

\begin{table*}[htbp]
    \vspace{-0.2cm}
    \centering
    \caption{RQ2: \toolname vs. NMT-based APRs with the perfect fault information provided on Defects4J V1.2}
    \resizebox{.7\linewidth}{!}{
    \begin{threeparttable}
        \begin{tabular}{lcccccccccc}
        \toprule
        \textbf{Projects} & \textbf{\toolname} & \textbf{ThinkRepair*} & \textbf{KNOD} & \textbf{TENURE} & \textbf{SelfAPR} & \textbf{RewardRepair} & \textbf{CURE} & \textbf{CoCoNuT} & \textbf{DLFix} & \textbf{DeepDebug}\\
        \midrule
        \textbf{Chart} & 
        \cellcolor{lightgray}11 & 6 & 9 & 6 & 7 & 5 & 9 & 6 & 5 & - \\
        \textbf{Closure} & \cellcolor{lightgray}31 & 19 & 22 & 21 & 16 & 15 & 13 & 8 & 10 & - \\
        \textbf{Lang} & \cellcolor{lightgray}19 & 16 & 11 & 12 & 10 & 7 & 9 & 7 & 7 & - \\
        \textbf{Math} & \cellcolor{lightgray}27 & 22 & 18 & 16 & 18 & 18 & 16 & 12 & 12 & - \\
        \textbf{Mockito} & \cellcolor{lightgray}6 & 5 & 5 & 3 & 2 & 2 & 4  & 4 & 1 & - \\
        \textbf{Time} & \cellcolor{lightgray}4 & 2 & 2 & 3 & 3 & 1 & 1 & 1 & 2 & - \\
        \midrule
        \textbf{\# Sum} & \cellcolor{lightgray}98 & 70 & 67 & 61 & 56 & 48 & 52 & 38 & 37 & - \\
        \bottomrule
        \end{tabular}%
        ``-'': indicates no results reported in the original work~\cite{drain2021deepdebug}.
    \end{threeparttable}
    } 
    \label{tab:vs_nmt}%
    \vspace{-0.3cm}
\end{table*}%

\begin{figure}[htbp]
    \centering
    \vspace{-0.3cm}
    \includegraphics[width=\linewidth]{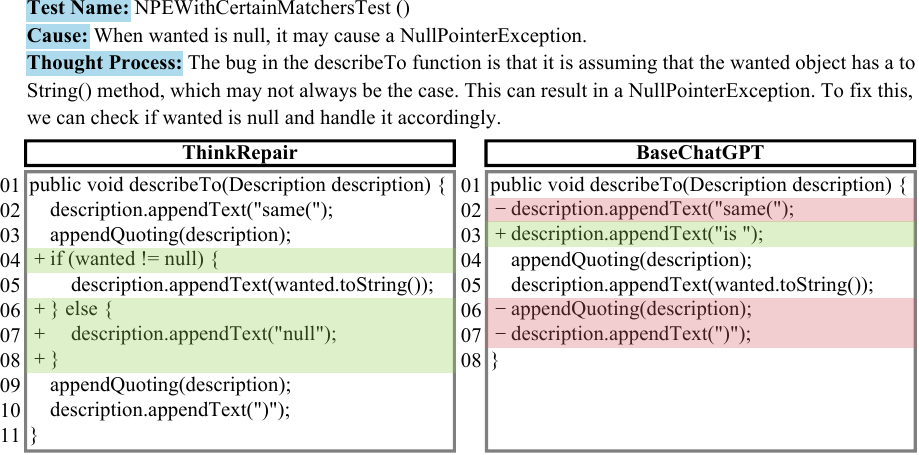}
    \caption{RQ1: Unique bug fixed by \toolname on Defects4J}
     \vspace{-0.3cm}
    \label{fig:unique_vs}
\end{figure}

To illustrate the effectiveness of \toolname compared to Base-ChatGPT, 
we present an instance of a bug (\href{https://github.com/mockito/mockito/issues/229}{Mockito-29}) in Defects4J V1.2 that is exclusively fixed by \toolname, as depicted in Fig.~\ref{fig:unique_vs}. 
The bug causes an incorrect description generation when the variable \textit{wanted} is null.
Both BaseChatGPT and \toolname offer distinct solutions to resolve this bug.
In this case, BaseChatGPT's solution mistakenly alters the line \textit{description.appendText("same(")} by substituting it with \textit{description.appendText("is\ ")}, causing a deviation from the intended code functionality.
Conversely, \toolname accurately recognizes the importance of preserving the original line and concentrates on enhancing other segments of the code.
It introduces additional conditional statements to handle scenarios where the \textit{wanted} variable could be null, which  guarantees proper behavior of the code when \textit{wanted} is not null and finally repairs the bug with the conservation of intended functionality.

\intuition{
{\bf Answer to RQ-1}: 
The Basic LLMs have limited ability to fix bugs and \toolname can improve it with suitable adaptions.
Overall, \toolname performs better than LLM-based APRs which indicates the priority by properly knowledge collection, combining few-shot selection as well as interaction feedback.
}

\subsection{RQ-2: Compare with NMT-based APRs}
\label{sec:rq2}

\noindent
{\bf \underline{RQ2-Analysis Procedure.}} 
{We compare \toolname with 8 NMT-based SOTA baselines: KNOD~\cite{jiang2023knod}, TENURE~\cite{meng2023tenure}, SelfAPR~\cite{ye2022selfapr}, RewardRepair~\cite{ye2022neural}, CURE~\cite{jiang2021cure}, DeepDebug~\cite{drain2021deepdebug}, CoCoNuT~\cite{lutellier2020coconut}, and DLFix~\cite{li2020dlfix}.
Benefiting from the powerful learning capability of deep neural networks, these NMT-based SOTAs have also been verified for their effectiveness on bug-fixing tasks.}

{\bf Data Splitting, Fault Information and Experiment Settings.} 
Like RQ-1, we use Defects4J and Quixbugs datasets and consider the same data splitting.
Following previous works~\cite{jiang2023knod, meng2023tenure, ye2022selfapr, ye2022neural, jiang2021cure, drain2021deepdebug, lutellier2020coconut, li2020dlfix}, we directly adopt the results from the original papers when the perfect fault information is provided. 
This comparison setting is the preferred or the only one for comparing recent NMT-based APRs as it removes the influence of other factors, such as fault localization, thus showing the pure potential of different approaches~\cite{xia2022less}.
We also focus on single function bug-fixing scenarios and use the same evaluation settings (e.g., \textit{Contrastive-based Selection} for few-shot selection and five interaction feedbacks).

\noindent
\underline{\textbf{RQ2-Results.}} 
Table~\ref{tab:vs_nmt} shows that \textbf{\toolname can outperform the studied 8 state-of-the-art NMT-based APR approaches on Defects4J V1.2}. 
With the perfect fault information, \toolname achieves the best overall performance with a total of 98 bug fixes, representing a significant improvement over the NMT-based baselines, in the range of 46.3\% (67 of KNOD)$\sim$164.9\% (37 of DLFix).
In addition, \toolname performs the best in all six projects with improvements ranging from 1 to 9 bugs, which indicates the superiority of our proposed approach.
On the open-source LLM (i.e., CodeLlama), ThinkRepair* outperforms supervised APRs.
Specifically, ThinkRepair* fixes 3$\sim$33 more bugs than the SOTA APRs.

Furthermore, \toolname generates far fewer patches ($\leq$ 125 in total per bug) than the NMT-based approaches that may generate up to 10,000 patches per bug ~\cite{jiang2021cure, lutellier2020coconut}, which indicates \toolname can also ensure a good efficiency.

\begin{figure}[htbp]
    \centering
    \vspace{-0.3cm}
    \includegraphics[width=.6\linewidth]{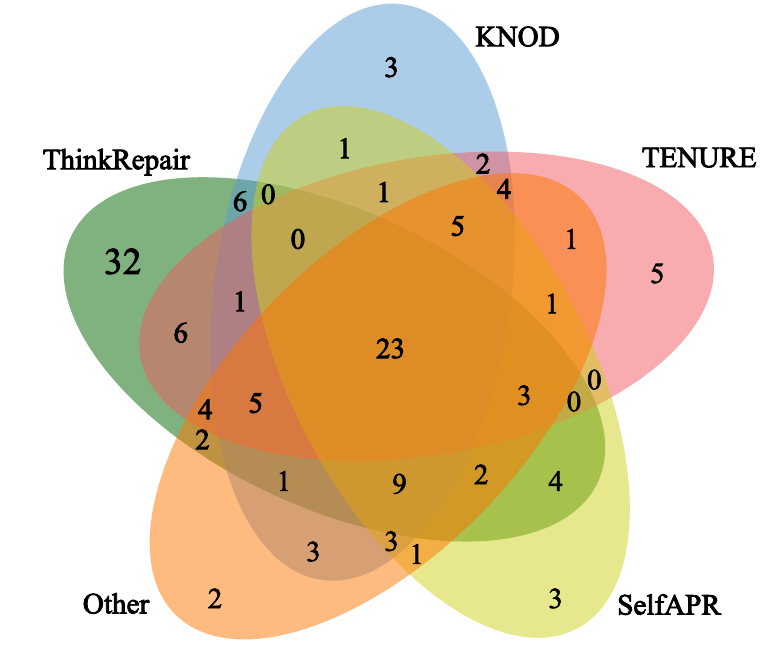}
    \caption{RQ2: Bug-fixing Venn diagram on Defects4J V1.2 of \toolname and all studied NMT-based APRs}
    \vspace{-0.3cm}
    \label{fig:venn}
\end{figure}

We draw a Venn diagram to further illustrate the performance difference on bug-fixing.
For a better presentation, we independently illustrate the Top-3 best baselines (i.e., KNOD, TENURE and SelfAPR) on the basis of the number of correctly fixed bugs and divide the rest methods into one group named ``Other'' for easy reference.
As for the ``Other'' group, we union all distinct correctly fixed bugs by the rest methods for comparison.
Fig.~\ref{fig:venn} shows the illustrated results and we can also obtain two observations: 
(1) Individual approaches have varying capabilities of fixing bugs and each of them can fix some specific bugs that other approaches cannot address.
Therefore, to some extent, these methods have a complementary performance.
(2) Overall, \toolname has a more powerful ability than baselines since it can auto-fix the most number of unique bugs (i.e., 32) that other baselines can hardly fix.

\begin{figure}[!htbp]
    \centering
    \vspace{-0.2cm}
    \includegraphics[width=\linewidth]{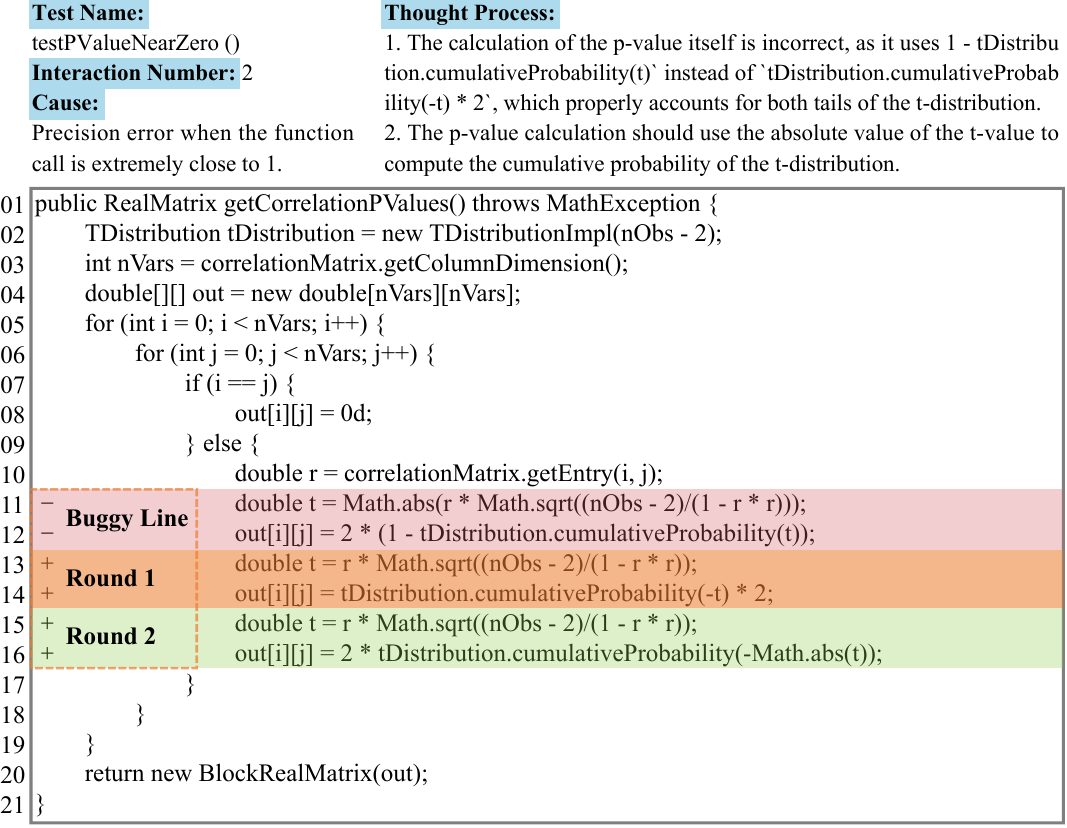}
    \caption{RQ2: Unique bug fixed in Defects4J V1.2}
    \vspace{-0.3cm}
    \label{fig:unique}
\end{figure}

{\bf Case Study.} To further understand why \toolname has an outstanding performance in fixing unique bugs, we further analyze one example (i.e., \href{https://issues.apache.org/jira/browse/MATH-371}{Math-69} in Defects4J V1.2) as a case study in Fig.~\ref{fig:unique}. 
The function calculates a matrix of $p$-values of a 2-sided, 2-sample $t$-test.
The bug is caused by a precision error when the function call is extremely close to 1. 
This is a hard-to-fix bug since the change is quite subtle and it does not fit any of the common templates used in traditional APRs.
To generate the correct patch, \toolname needs to understand the goal of the function (i.e., $p$-value calculation) and correctly use statistical formulas. 
As shown in Fig.~\ref{fig:unique}, \toolname undergoes two attempts (i.e., marked as \textit{``Round 1''} and \textit{``Round 2''}), and the \textit{``Thought Process''} reveals how \toolname thinks to solve this bug.
Specifically, the first attempt (i.e., Lines 13-14) modifies the buggy lines (i.e., Lines 11-12) to \textit{``tDistribution.cumulativeProbability(-t) * 2''} while overlooking the necessity of using the absolute value of the \textit{t-value}.
In the second attempt (i.e., Lines 15-16), \toolname comprehends the semantics gleaned from the test failure message and thereby successfully fixes the bug. 
This example further exemplifies the capability of \toolname to leverage previously overlooked semantic information within failing tests to directly guide the repair process.

Table~\ref{tab:vs_nmt_additional} shows that compared with the NMT-based APRs, \toolname can achieve the best performance on Defects4J V2.0, QuixBugs-Java, and QuixBugs-Python, similar to the findings obtained Defects4J V1.2. 
\toolname can auto-fix 107 bugs on Defects4J V2.0, with 60$\sim$65 (127.7\%$\sim$154.8\%) more bugs than the current state-of-the-art NMT-based APRs.
Furthermore, ThinkRepair* fixes 72 bugs and improves baselines by 53.2\%$\sim$71.4\%.
Additionally, \toolname has improved NMT-based baselines by fixing 14$\sim$26 more bugs on QuixBugs-Java and 19$\sim$21 bugs on QuixBugs-Python.

\begin{table}[!htbp]
    \centering
    \vspace{-0.2cm}
    \caption{RQ2: \toolname vs. NMT-based APRs with perfect fault information provided on the other three datasets}
    \resizebox{.8\linewidth}{!}{
    \begin{threeparttable}
        \begin{tabular}{lcccc}
        \toprule
        \multirow{2}[1]{*}{\textbf{Models}} & \textbf{Defects4J} & \multicolumn{2}{c}{\textbf{QuixBugs}} \\
        \cmidrule{3-4}& \textbf{V2.0 (228 bugs)} & \textbf{Java (40 bugs)} & \textbf{Python (40 bugs)}\\
        \midrule
        KNOD & 47 & 25 & - \\
        TENURE & 43 & - & \\
        SelfAPR & 42 & - & - \\
        RewardRepair & 44 & 20 & - \\
        CURE & - & 21 & - \\
        DeepDebug & - & - & 21 \\
        CoCoNuT & - & 13 & 19 \\
        DLFix & - & - & - \\
        \midrule
        ThinkRepair* & 72 & 38 & 38 \\
        \rowcolor{lightgray}  \toolname & 107 & 39 & 40 \\
        \bottomrule
        \end{tabular}
        ``-'': indicates no results reported in the original work.
        \end{threeparttable}
    }
    \vspace{-0.4cm}
    \label{tab:vs_nmt_additional}
\end{table}

\intuition{
{\bf Answer to RQ-2}:
All APR approaches have complementary advantages in fixing different bugs.
Overall, \toolname outperforms NMT-based approaches in terms of the number of auto-fixed bugs.}

\subsection{RQ-3: Configurations of \toolname}
\label{sec:rq3}

\noindent
{\bf \underline{RQ3-Analysis Procedure.}} 
\toolname has two important components: \ding{172} \textbf{CoT few-shot learning (knowledge collection + few-shot selection)} prompts LLM to construct a chain of thought pool and selects diverse and effective examples for LLM to better understand the downstream task with few-shot learning, and \ding{173} \textbf{interaction feedback} provides feedback to LLM with the test failure information interactively during the process of function verification. 
Therefore, in this RQ, we aim to conduct a comprehensive experiment to evaluate the impact of different components on the \toolname's performance.
Furthermore, we study the impact of the number of interactions with LLM and the impact of the four few-shot selection strategies on the performance of \toolname.

{\bf Data Splitting and Fault Information.}
In this RQ, we utilized ChatGPT as the backend LLM.
Considering the cost of invoking the ChatGPT API multiple times for this ablation study, we concentrate our experiments on Defects4J V1.2, which has the most number of single function bugs (i.e., 255) among our studied datasets.
Also, we only provide \toolname with the method-level fault information but no statement-level fault information.

{\bf Experiment Settings.}
As for the interaction number, the maximum interaction number (cf. Section~\ref{sec:fixing}) dictates the amount of history/feedback within each individual repair query.
Therefore, when interacting once, it is equivalent to directly using the initial prompt without any feedback for \toolname. 
We treat the original ChatGPT as the basic model for comparison and investigate its initial performance in the zero-shot learning setting and few-shot learning setting.
Zero-Shot means that we directly prompt ChatGPT to fix a bug without providing any other information.
Few-Shot means that we provide ChatGPT with two examples of fixes but do not include CoT information (i.e., directly using existing bug-fixing data as examples).
Moreover, \textit{``knowledge collection''} is a preparation step for \textit{``few-shot selection''} when fixing bugs.
Therefore, the two parts are used together.
We build a variant of \toolname without interaction feedback as \toolname-v1 and another variant without CoT few-shot learning as \toolname-v2. 
All of our ablation experiments utilize the default settings outlined in Section~\ref{sec:setting}.

\noindent
{\bf \underline{RQ3-Results.}} 
{\bf Impacts of Components.} 
According to the results in Table~\ref{tab:component}, we can observe that:
(1) In few-shot, ChatGPT exhibits better bug-fixing performance compared to zero-shot, but the performance improvement is limited (i.e., 34$\rightarrow$36).
(2) Two components have their own advantages in a single-function bug-fixing scenario, achieving a varying performance and significantly improving the performance of ChatGPT. 
Both of them can contribute to the performance of \toolname. 
(3) The combination between \textit{``Knowledge Collection''} and \textit{``Few-Shot Selection''} is effective for \toolname, which improves ChatGPT a lot (i.e., 36$\rightarrow$57) and emphasizes the importance of prompting ChatGPT by selecting high-quality of examples.
(4) The \textit{``Interaction Feedback''} seems to contribute the performance of \toolname, which brings ChatGPT with a large improvement (i.e., 36$\rightarrow$62).
It indicates that the information obtained from test failures can help ChatGPT understand the reasons for failures and provide guidance for generating plausible fixes.
(5) A combination of these two components can improve ChatGPT. 
\toolname is capable of generating 44 more correct fixes than ChatGPT with a few-shot setting (i.e., 36$\rightarrow$80).

\begin{table}
    \centering
    \vspace{-0.2cm}
    \caption{RQ3: The performance difference among different components}
    \resizebox{.8\linewidth}{!}
    {
        \begin{tabular}{lcccr}
        \toprule
        \textbf{Models} & \textbf{\makecell{CoT Few-Shot \\Learning}} &  \textbf{\makecell{Interaction\\Feedback}} & \textbf{\makecell{\# Correct\\Fixes}} \\
        \midrule
        Zero-Shot & \ding{55} & \ding{55} & 34 \\
        Few-Shot & \ding{55} & \ding{55} & 36 \\
        \midrule
        \toolname-v1  & \ding{51} & \ding{55} & 57 \\
        \toolname-v2  & \ding{55} & \ding{51} & 62 \\
        \midrule
        \rowcolor{lightgray} {\toolname}  & \ding{51} & \ding{51} & {80} \\
        \bottomrule
        \end{tabular}
    }
    \vspace{-0.2cm}
    \label{tab:component}
\end{table}

\begin{figure}[htbp]
    \vspace{-0.2cm}
    \centering
    \includegraphics[width=.8\linewidth]{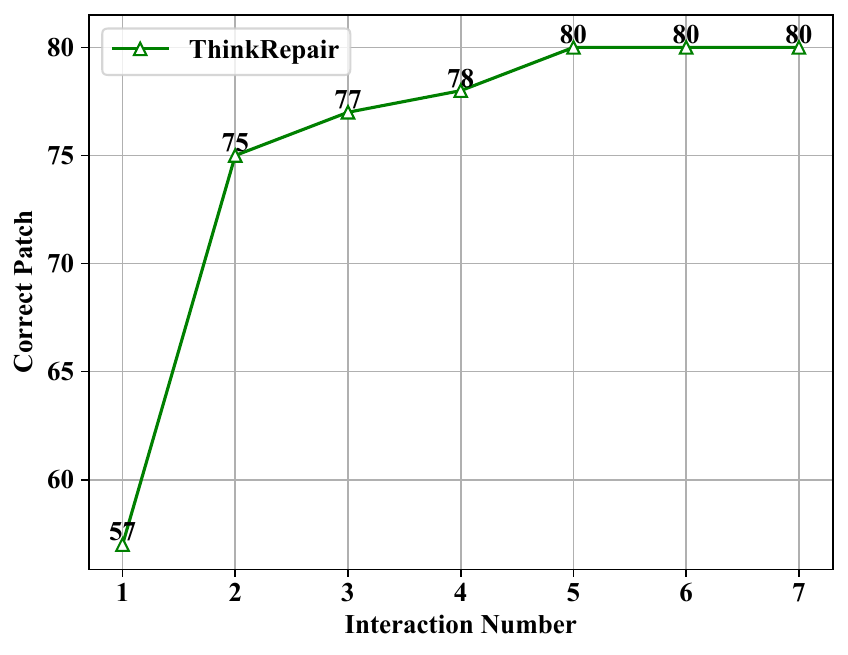}
    \caption{\xy{RQ3: The varying performance of \toolname with different interaction number}}
    \label{fig:interaction}
    \vspace{-0.3cm}
\end{figure}

\textbf{{Impacts of Interaction Number.}}
According to the results in Fig.~\ref{fig:interaction}, we observe that: 
(1) Different interaction numbers have varying impacts on \toolname's performance and the performance of \toolname increases as the number of interactions increases.
(2) Sampling directly from the ChatGPT (i.e., interact once) may not ensure a good performance.
For example, when directly interacting with ChatGPT without information feedback, \toolname achieves the lowest number of correct fixes (i.e., 57). 
(3) Feedback information promotes ChatGPT to reason, but more interaction times may not guarantee additional performance improvement.
Notice that \toolname has a big improvement when interacting twice with ChatGPT (i.e., 57$\rightarrow$75). 
However, when continuously increasing the number of interactions, the rate of performance improvement decreases (i.e., 75$\rightarrow$80) and meanwhile, the interaction cost with ChatGPT is increasing.
Considering both the performance improvement and the communication cost caused by ChatGPT, we adopt five times of interactions as the default setting.

\begin{table}[!htbp]
    \centering
    \vspace{-0.2cm}
    \caption{\xy{RQ3: The performance difference among four different few-shot selection strategies}}
    \resizebox{.8\linewidth}{!}
    {
    \begin{threeparttable}   
    
        \begin{tabular}{lrrrr}
        \toprule
        \textbf{Datasets} & \textbf{CSelect} & \textbf{SSelect} & \xy{\textbf{ISelect}} & \textbf{RSelect} \\
        \midrule
        {Defects4J V1.2 (255 bugs)} & \textbf{80} & 73 & \xy{70} & 68 \\
        {Defects4J V2.0 (228 bugs)} & \textbf{90} & 84 & \xy{78} & 77 \\
        {QuixBugs-Java (40 bugs)} & \textbf{39} & 38 & \xy{38} & 38 \\
        {QuixBugs-Python (40 bugs)} & \textbf{40} & \textbf{40} & \xy{39} & 39 \\
        \midrule
        \textbf{\# Sum} & \textbf{249}& 235 & \xy{225} & 222 \\
        \bottomrule
        \end{tabular}
        ``CSelect'': Contrastive-based Selection, ``SSelect'': Semantic-based Selection, 
        \xy{``ISelect'': IR-based Selection}, 
        ``RSelect'': Randomly Selection.
        \end{threeparttable}
    }
    \label{tab:selection}
\end{table}

\xy
{
\textbf{{Impacts of Few-Shot Selection Strategy.}}
According to the results in Table~\ref{tab:selection}, we observe that: 
(1) Both \textit{Contrastive-based Selection} and \textit{Semantic-based Selection} help to generate more correct fixes than \textit{IR-based Selection} and \textit{Randomly Selection}, which indicates the importance of the semantic similarity among samples. Particularly, \textit{Contrastive-based Selection} and \textit{Semantic-based Selection} can help \toolname fix 27 and 13 more bugs than does \textit{Randomly Selection} help, respectively.
(2) Overall, \textit{Contrastive-based Selection} performs the best and shows promising results in selecting high-quality examples, which may benefit from the well-trained semantic encoder (i.e., UniXcoder) fine-tuned with contrastive learning.
}

\xy
{
We also discuss the cost of different selection strategies. 
It has some difference in time cost.
On average, both \textit{Semantic-based Selection} and \textit{Contrastive-based Selection} strategies take about 0.01 seconds to embed a function, while \textit{IR-based Selection} and \textit{Randomly Selection} strategies take 0.9 milliseconds. 
However, considering that a repair process is mainly consumed in the output of the LLM (which usually takes several seconds), the time cost of selection is negligible. 
Therefore, to strike a balance between efficiency and effectiveness, we adopt \textit{Contrastive-based Selection} as default.
}

\intuition{{\bf Answer to RQ-3}: 
(1) The two components (i.e., \textit{CoT few-shot learning and interaction feedback}) contribute substantially to \toolname, and combining them achieves the best performance.
(2) Increasing the interaction number (e.g., from 1 to 5) can significantly improve the performance of \toolname, but more interactions may not gain larger performance improvement.
(3) Semantically similarity-based example selection strategies can pick out high-quality examples for \toolname than random selection and \textit{Contrastive-based Selection} is the best choice.
}
\label{sec:results}

\section{Discussion}

\xy
{
This section discusses open questions regarding the data leakage and threads to the validity of \toolname.
}

\subsection{Evaluation of Data Leakage}

\subsubsection{Similarity between generated patches and the ground-truth}
\label{sec:data_leakage_1}

\xy
{
We follow the work~\cite{xia2022less} and initially calculate the number of correct patches, 
which lexically matches the ground-truth in Defects4J V1.2.
We find that out of 98 correct patches, 24 of them lexically match the ground-truth (24.5\%), while the other patches semantically match the ground-truth.
We exemplify with Fig~\ref{fig:unique},
the ground-truth fix is \textit{``double t = Math.abs(r * Math.sqrt((nObs - 2)/(1 - r * r))); out[i][j] = 2 * tDistribution.cumulativeProbability(-t);''}. 
Our \toolname generates a fix that is different from the ground-truth but semantically equivalent, namely \textit{``double t = r * Math.sqrt((nObs - 2)/(1 - r * r)); out[i][j] = 2 * tDistribution.cumulativeProbability(-Math.abs(t));''}.
}

\subsubsection{Study on Real-World Projects}
\label{sec:data_leakage_2}

\xy
{
We follow Defects4J and collect bug-fixing commits from high-quality open-source projects included in Defects4J.
The resulting dataset is referred to as RWB (Real-World Bugs).
We evaluate data leakage on both ChatGPT and DeepSeek, denoting them as \underline{\toolname and ThinkRepair*}, respectively.
Notice that the pre-training data for ChatGPT was collected before September 2021~\cite{openai2022chatgpt}, while the pre-training data for DeepSeek was collected from GitHub before February 2023~\cite{deepseek-coder}. 
Consequently, we collected two datasets to assess data leakage. 
The first dataset (RWB V1.0) comprises bug-fixing commits after October 2021, while the second dataset (RWB V2.0) includes bug-fixing commits after March 2023, resulting 113 and 61 bugs, respectively.
}

\xy
{
Since \toolname focuses on single-function bugs, we perform two steps on the original datasets to obtain valid functions:
}

\xy
{
\textbf{Step-1:} Each commit is considered a mini-version of a project.
We use the commit IDs to request commit histories of the projects, and for each commit, we extract the code changes between before and after fixing a bug.
Finally, we use the code change information to obtain the buggy and fixed version of a function. 
}

\xy
{
\textbf{Step-2:} To clean and normalize the dataset, we keep only single-function bugs.
In this step, we finally obtain the RWB V1.0 dataset, which comprises 44 single-function bugs (4 of Cli, 5 of Codec, 1 of Collections, 8 of Compress, 2 of Csv, 6 of Lang, and 18 of Jsoup), and the RWB V2.0 dataset, which comprises 29 single-function bugs (4 of Cli, 4 of Codec, 4 of Compress, 6 of Lang, and 11 of Jsoup).
}

\xy
{
Finally, we conducted a comparative analysis on \toolname and AlphaRepair with perfect fault information, and AlphaRepair was identified as the baseline with the best performance (cf. Section~\ref{sec:rq1}).
Table~\ref{tab:data_leakage} presents the results of \toolname on RWB. 
According to the results, we can conclude that \toolname possesses the ability to fix bugs in real-world projects, not just limited to the bugs present in the Defects4J dataset and QuixBugs dataset. 
Furthermore, both \toolname and ThinkRepair* outperform AlphaRepair, underscoring the practicality and viability of our approach.
}

\xy
{
In summary, we believe that data leakage will not significantly affect the performance of our \toolname.
}

\begin{table}[htbp]
  \centering
  \caption{\xy{\toolname vs. AlphaRepair on RWB}}
  \resizebox{\linewidth}{!}
  {
  \begin{threeparttable}
    \begin{tabular}{l|cc|cc}
    \toprule
    \multicolumn{1}{l|}{\multirow{2}[2]{*}{\textbf{Projects}}} & \multicolumn{2}{c|}{\textbf{RWB V1.0 (44 bugs)}} & \multicolumn{2}{c}{\textbf{RWB V2.0 (29 bugs)}} \\
    \cmidrule{2-5} 
    \multicolumn{1}{l|}{} & \textbf{ThinkRepair} & \textbf{AlphaRepair}  & \textbf{ThinkRepair*} &  \textbf{AlphaRepair} \\
    \midrule
    \textbf{Cli} & \cellcolor{lightgray}4 & 3 & \cellcolor{lightgray}4 & 3 \\
    \textbf{Codec} & \cellcolor{lightgray}3 & 1 & \cellcolor{lightgray}1 & 1 \\
    \textbf{Collections} & \cellcolor{lightgray}1 & 0 & \cellcolor{lightgray}- & - \\
    \textbf{Compress} & \cellcolor{lightgray}1 & 1 & \cellcolor{lightgray}- & - \\
    \textbf{Csv} & \cellcolor{lightgray}1 & 0 & \cellcolor{lightgray}- & - \\
    \textbf{Lang} & \cellcolor{lightgray}3 & 2 & \cellcolor{lightgray}3 & 1 \\
    \textbf{Jsoup} & \cellcolor{lightgray}6 & 2 & \cellcolor{lightgray}2 & 1 \\
    \midrule
    \textbf{\# Sum} & \cellcolor{lightgray}19 & 9 & \cellcolor{lightgray}10 & 6 \\
    \bottomrule
    \end{tabular}%
    \end{threeparttable}
    } 
    \vspace{-0.3cm}
  \label{tab:data_leakage}%
\end{table}%

\subsection{Threats to Validity}

\noindent
\textbf{Internal Validity.}
The first internal threat arises from the manual validation employed to determine the correctness of the plausible patches.
To mitigate this concern, we conduct a thorough examination and comprehensive discussion of each patch, following prior work~\cite{jiang2021cure, xia2023automated, xia2022less, ye2022selfapr, ye2022neural, zhu2021syntax}.
The second one comes from potential data leakage since referenced developer patches may be part of the training data of LLM.
\xy
{
As discussed in section~\ref{sec:data_leakage_1}, 24.5\% of correct patches aligned with the reference developer fix. 
Moreover, even after excluding all the correct patches (24) that aligned with the reference developer patch, \toolname is still capable of generating the correct patch for 29 unique bugs, none of which could be fixed by any previous methods.
Additionally, compared to BaseChatGPT, \toolname achieves 46 more correct fixes.
This demonstrates that the improved results achieved by \toolname are not merely a result of memorizing the training data.
We also collected bugs from real-world projects to evaluate data leakage. 
\toolname can fix 19 out of 44 bugs on RWB V1.0, and 10 out of 29 bugs on RWB V2.0.
}

\noindent
\textbf{External Validity.}
The effectiveness observed in \toolname's performance may not be applicable across different repair datasets. 
We conduct evaluations not only on the widely-used Defects4J dataset but also on QuixBugs dataset. 
This broader evaluation scope aims to showcase the generalizability of our approach.


\label{sec:discussion}

\section{Related Work}
\subsection{Large Language Model}

Large Language Models (LLMs)~\cite{brown2020language} have been widely adopted since the advances in Natural Language Processing (NLP) which enable LLM to be well-trained with both billions of parameters and billions of training samples, and consequently, they bring a large performance improvement.
LLMs can be easily used for a downstream task by being fine-tuned~\cite{radford2018improving} or being prompted~\cite{liu2023pre} since they are trained to be general and they can capture different knowledge from various domain data.
Fine-tuning is used to update model parameters for a particular downstream task by iterating the model on a specific dataset.
Meanwhile, prompting can also be directly used by providing natural language descriptions or a few examples of the downstream task.
Compared to prompting, fine-tuning is expensive since it requires additional model training and has limited usage scenarios, especially in cases where sufficient training datasets are unavailable.

ChatGPT~\cite{openai2022chatgpt} is a successor of InstructGPT~\cite{ouyang2022training} and is fine-tuned with the Reinforcement Learning with Human Feedback (RLHF) approach~\cite{christiano2017deep,ouyang2022training,ziegler2019fine}.
RLHF first fine-tunes the model with the input (i.e., a small dataset of prompts) and the desired output (i.e., usually human-written).
Following that, a reward model will be trained on a larger set of prompts by sampling a few outputs that are generated by the fine-tuned model and these outputs are re-ordered by humans.
Finally, reinforcement learning~\cite{schulman2017proximal} is adopted to calculate the reward of each output that is generated based on the reward model and eventually updates the LLM parameters accordingly.
Benefiting from fine-tuning as well as human preference alignments, LLM has a better understanding of input prompts and instructions to perform better on various downstream tasks~\cite{bang2023multitask,ouyang2022training}.

\subsection{Automated Program Repair}

Automated Program Repair (APR) can assist developers in generating patches for specific bugs based on their potential fault locations. 
Traditional APR techniques can be classified into heuristic-based~\cite{le2016history, le2011genprog, wen2018context}, constraint-based~\cite{demarco2014automatic, le2017s3, long2015staged, mechtaev2016angelix}  and template-based~\cite{liu2019tbar, ghanbari2019practical, hua2018sketchfix, liu2019avatar, martinez2016astor} approaches.
Template-based APR tools have gained recognition as state-of-the-art due to their ability to fix a large number of bugs. 
These tools utilize human-defined or automatically-mined templates to identify potential buggy code patterns and apply corresponding fixes.
However, template-based tools are limited to the patterns within their predefined set and lack the ability to generalize to other types of bugs or fixes. 
To address this limitation, researchers have proposed learning-based APR techniques by leveraging recent advancements in Deep Learning.
Techniques based on Neural Machine Translation (NMT) have been extensively studied in recent years~\cite{ye2022selfapr, ye2022neural, zhu2021syntax, jiang2021cure, lutellier2020coconut, jiang2023knod, meng2023tenure, drain2021deepdebug, li2020dlfix,li2022dear,li2020improving} and they treated APR as an NMT problem that is translating buggy code into correct code. 
However, these methods heavily rely on historical bug-fixing data and noise may impact their performance.

In order to overcome the limitations of NMT-based tools, researchers have explored the potential of utilizing LLMs directly to generate correct patches. 
By pre-training on large amounts of open-source code snippets, LLMs have the ability to generate correct code directly based on the surrounding context, eliminating the need for translation from the buggy code.  
AlphaRepair~\cite{xia2022less} is the first tool for cloze-style APR and its performance indicates that LLM-based APR outperforms the widely studied NMT-based APR techniques in real-world systems. 
Following that, researchers~\cite{prenner2022can,kolak2022patch} directly adopt Codex to generate a fixed function based on a buggy one.
Recently, Xia et al.~\cite{xia2023automated} conducted an extensive study of LLM-based APR techniques using various LLMs~\cite{chen2021evaluating,black2022gpt,wang2021codet5,fried2022incoder} and their results further demonstrated the superiority of LLM-based APR.
\xy
{
ChatRepair~\cite{xia2023keep} simply uses the chat capability of ChatGPT and iteratively enters test information to obtain the final patch. 
However, the capabilities of LLMs are influenced by high-quality prompts.
In this paper, we propose a self-directed framework, \toolname.
The objective of \toolname is to enable the LLM to engage in self-reflection to construct a high-quality knowledge pool, and select few-shot examples from the knowledge pool to better guide LLM. 
}
\label{sec:related_work}

\vspace{-0.1cm}
\section{Conclusion}

This paper proposes a novel approach \toolname, which is a single function APR tool. 
\toolname has two main phases: the collection phase and the fixing phase.
The former phase adopts knowledge collection to generate a series of thought processes that provide high-quality examples for subsequent phases.
The latter phase targets fixing a bug by first selecting examples for few-shot learning and second automatically interacting with LLM, optionally appending with
feedback of testing information.
Through this repair paradigm, \toolname has strong analytical and reasoning abilities and is capable enough to repair complex bugs.
Therefore, \toolname achieves state-of-the-art performance on both Defects4J V1.2 and Defects4J V2.0, surpassing the baselines by 17$\sim$62 and 12$\sim$65 more bugs, respectively.
\label{sec:conclusion}

\section{Data Availability}
The replication of this paper is publicly available~\cite{replication}.

\section*{Acknowledgements}{
This work was supported by the National Natural Science Foundation of China (Grant No.62202419), the Fundamental Research Funds for the Central Universities (No. 226-2022-00064),
Zhejiang Provincial Natural Science Foundation of China (No. LY24F020008),
the Ningbo Natural Science Foundation (No. 2022J184), 
and the State Street Zhejiang University Technology Center.
}

\balance
\bibliographystyle{ACM-Reference-Format}
\bibliography{main}

\end{document}